\documentclass{vgtc}                          % final (conference style)

\graphicspath{{figures/}{pictures/}{images/}{./}}

\usepackage{times}                     
\usepackage{subcaption}

\vgtcinsertpkg

\usepackage{hyperref}
\usepackage{enumitem}
\usepackage{tikz}
\usepackage{xcolor}
\usepackage{textcomp}
\usepackage{latexsym}

\newlist{designNum}{enumerate}{1}
\setlist[designNum]{label=\bfseries D\arabic*., ref=\arabic*}

\definecolor{blueframe}{RGB}{146, 197, 222}
\definecolor{orangeframe}{RGB}{253, 145, 64}

\newcommand\bluecircle{%
  \begin{tikzpicture}[baseline=0.1ex,every node/.style={inner sep=0,outer sep=0}]
    \filldraw[draw=blueframe,fill=blueframe, thin, line width=1pt] (0.1,0.125) circle (0.1);
  \end{tikzpicture}%
}

\newcommand\orangecircle{%
  \begin{tikzpicture}[baseline=0.1ex,every node/.style={inner sep=0,outer sep=0}]
    \filldraw[draw=orangeframe, fill=orangeframe, thin, line width=1pt] (0.1,0.125) circle (0.1); 
  \end{tikzpicture}%
}

\title{Testing the Test: Observations When Assessing\\ Visualization Literacy of Domain Experts}

\author{Seyda Öney\thanks{e-mail: firstname.surname@visus.uni-stuttgart.de}\\ %
        \scriptsize University of Stuttgart %
\and Moataz Abdelaal\footnotemark[1]\\ %
        \scriptsize University of Stuttgart %
\and Kuno Kurzhals\footnotemark[1]\\ %
        \scriptsize University of Stuttgart %
\and Paul Betz\thanks{e-mail: firstname.surname@sowi.uni-stuttgart.de}\\ %
        \scriptsize University of Stuttgart %
\and Cordula Kropp\footnotemark[2]\\ %
        \scriptsize University of Stuttgart %
\and Daniel Weiskopf\footnotemark[1]\\ %
        \scriptsize University of Stuttgart %
        }

\abstract{
Various standardized tests exist that assess individuals' visualization literacy. Their use can help to draw conclusions from studies. However, it is not taken into account that the test itself can create a pressure situation where participants might fear being exposed and assessed negatively. This is especially problematic when testing domain experts in design studies. We conducted interviews with experts from different domains performing the Mini-VLAT test for visualization literacy to identify potential problems. Our participants reported that the time limit per question, ambiguities in the questions and visualizations, and missing steps in the test procedure mainly had an impact on their performance and content. We discuss possible changes to the test design to address these issues and how such assessment methods could be integrated into existing evaluation procedures.
}


\begin{document}

\maketitle

\section{Introduction}
 
Visualization literacy is important---not only to be able to \textit{``make meaning from and interpret patterns, trends, and correlations in visual representations of data''}~\cite{borner_investigating_2016} but also to be able to discern visualization mirages~\cite{Andrew2020Mirages}, misleading charts~\cite{LisnicPLK23}, and VisLies~\cite{vislies:online} that comes with the widespread use of social media~\cite{HuLWWSM12}.
Recently, the visualization community has focused 
on developing tools and instruments to measure visualization literacy~\cite{cui_adaptive_2024,ge_calvi_2023,lee_vlat_2017,pandey_minivlat_2023}. One example is Mini-VLAT, designed as a fast and effective way of measuring literacy among the general public~\cite{pandey_minivlat_2023}. With the goal of making the test practical and time-effective, the authors adopted twelve questions from the Visualization Literacy Assessment Test (VLAT)~\cite{lee_vlat_2017} and allocated 25 seconds for each question. 
While both VLAT and Mini-VLAT were designed to assess the literacy of novice users, in this interdisciplinary work with visualization experts and social scientists, we investigate the applicability and usefulness of Mini-VLAT when working with domain experts. In particular, we conducted structured interviews with seven experts from social science, architecture, engineering, and computer science backgrounds to obtain qualitative insights into the structure, types of questions, and sources of confusion or ambiguities encountered during the test.

The results show that the concept of a timed ``test'' may have introduced other factors such as stress and time pressure, suggesting that those who performed well on the test were likely better at managing stress and time constraints rather than inherently better at interpreting or reading visualizations. These factors might be undesirable when Mini-VLAT is used in the context of evaluating visualization tools with domain experts, where the experts' ability to work quickly under pressure is not relevant to the evaluation. In addition, we identified issues with test questions and ideas on how questions could be adapted to different domains. 

The contributions of this work include: (1) obtaining insights into the visualization literacy of domain experts, (2)
identifying issues in the Mini-VLAT test when working with domain experts, (3)~suggestions for improvements, and guidelines for using the test in different evaluation settings.

\section{Related Work}

\begin{figure*}[t]
    \centering   \includegraphics[width=\linewidth]{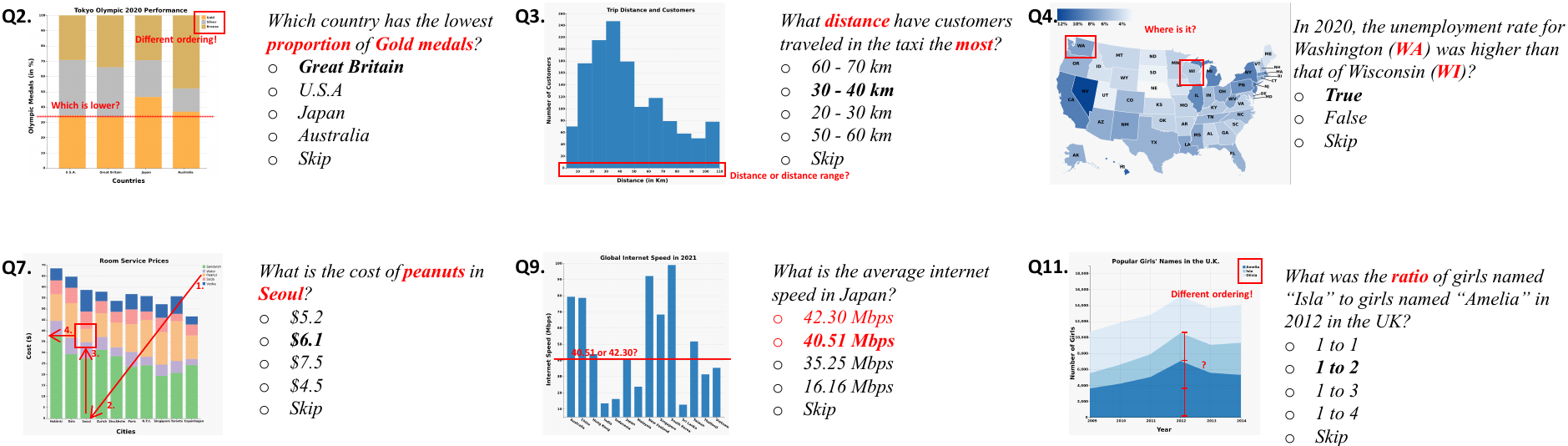}
    \caption{Six questions from Mini-VLAT with which participants had difficulties. The issues in the visualizations and questions are highlighted in red. Visualizations reprinted with permission from Pandey and Ottley~\cite{pandey_minivlat_2023} {\textcopyright} John Wiley \& Sons, Inc., and adapted.}
    \label{fig:mini_vlat_questions_issues}
\end{figure*}

While the research literature on visualization literacy 
has been growing over the past few years~\cite{firat_interactive_2022}, Taylor~\cite{taylor2003new} traced the term ``visual literacy'' back to the 1950s when photography was growing as a medium of communication. In his exploratory work, Taylor examined the different uses and misuses of the term ``literacy'' and coined the term ``visual information literacy,'' which is the closest to what the community refers to today as ``visualization literacy.''  

In psychometric test theory~\cite{cohen1996psychological, erguven2013two,mcdonald2013test}, there are different methods of assessing knowledge. In the pioneering work~\cite{boy_principled_2014}, Boy et al.\ proposed the Item Response Theory~\cite{van2018handbook}.
Lee et al.~\cite{lee_vlat_2017} also used this method to build VLAT. As part of test development, both validity and reliability were evaluated. The reliability coefficient omega~\cite{mcdonald2013test} measured the consistency of the test scores, while the content validity ratio~\cite{lawshe1975validitiy} assessed the appropriateness of individual items. Similarly, the quality of empirical studies can be determined by different types of validation, including construct, internal, external, and statistical validity~\cite{Carpendale2008}. 

The VLAT test contains 12 chart types and 53 multiple-choice questions, each with a 25-second limit. 
Motivated by the widespread occurrence of misinformation and misleading charts, Ge et al.~\cite{ge_calvi_2023} developed CALVI, a 45-item literacy test to assess the ability of people to reason about eleven common chart misleaders. 
While both VLAT and CALVI showed a high reliability score, both tests are time-consuming and require high cognitive efforts, which limits their usage and makes them hard to incorporate, especially if they were to be used as a part of a larger user evaluation.
To address this issue, Cui et al.~\cite{cui_adaptive_2024} employed computerized adaptive testing (CAT) to select the test items based on the performance of the participants on the previously answered questions, leading to adaptive VLAT (A-VLAT) and adaptive CALVI (A-CALVI). 
Aiming for the same goal, Pandey and Ottley~\cite{pandey_minivlat_2023} obtained a shorter version of VLAT by selecting one question for each chart type based on the total-item correlation, resulting in Mini-VLAT. This 12-question literacy test can be completed in five minutes. 

While most of the previous work targeted crowd-sourced workers,
few attempted to assess the literacy of the general audience. In their museum study, Börner et al.~\cite{borner_investigating_2016} assessed the familiarity of 273 youth and adult museum visitors in the United States with 20 different visualization types. 
The results were interesting as they showed low literacy levels among participants, revealing a gap between what the academic circles deem as a common visualization and what the general audience is actually able to read and interpret. 
In the same vein, Peck et al.~\cite{peck2019data} found that the perception of visualizations is influenced by the personal experiences of 42 residents of rural Pennsylvania.

In this work, we are not developing or introducing a new literacy assessment method; rather, we are using the already existing methods to obtain insights into the literacy of specific demographics. Our paper is closely related to the work of Nobre et al.~\cite{NobreReading24}  and Lee et al.~\cite{LeeKHLKY16}. The former attempted to identify where and how 120 crowd-sourced workers make mistakes when they attempt the VLAT test, while the latter attempted to qualitatively understand how 13 university participants make sense of unfamiliar visualizations. While we share the same goal of obtaining qualitative insights into the structure, types of questions, and sources of ambiguities in the Mini-VLAT test, we differ in our work by focusing on the notion of time pressure introduced by having a 25-second limit and by targeting 
demographics that are underrepresented in crowd-sourced platforms. That is, experts from social science, engineering, architecture, and computer science. The latter aspect makes our work share similar characteristics with Peck et al.~\cite{peck2019data} and Börner et al.~\cite{borner_investigating_2016}.

\section{Preliminary Observations}\label{sec:preliminaryobservation}

\begin{figure*}[t]
    \centering
    \begin{subfigure}{0.45\linewidth}
    \centering
       \includegraphics[width=\linewidth]{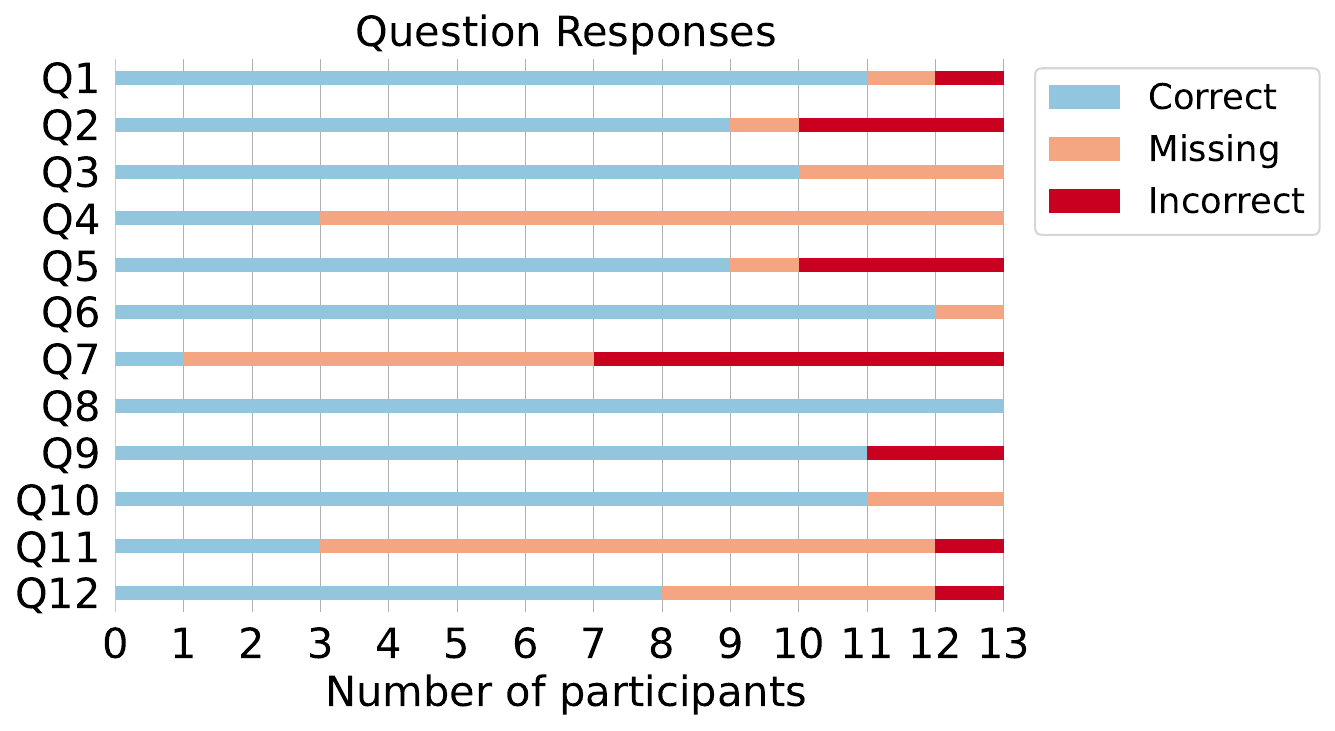}
    \caption{Responses}
    \label{fig:study_mini_vlat_result_responses}
    \end{subfigure}
    \hfill
    \begin{subfigure}{0.45\linewidth}
    \centering
    \includegraphics[width=\linewidth]    {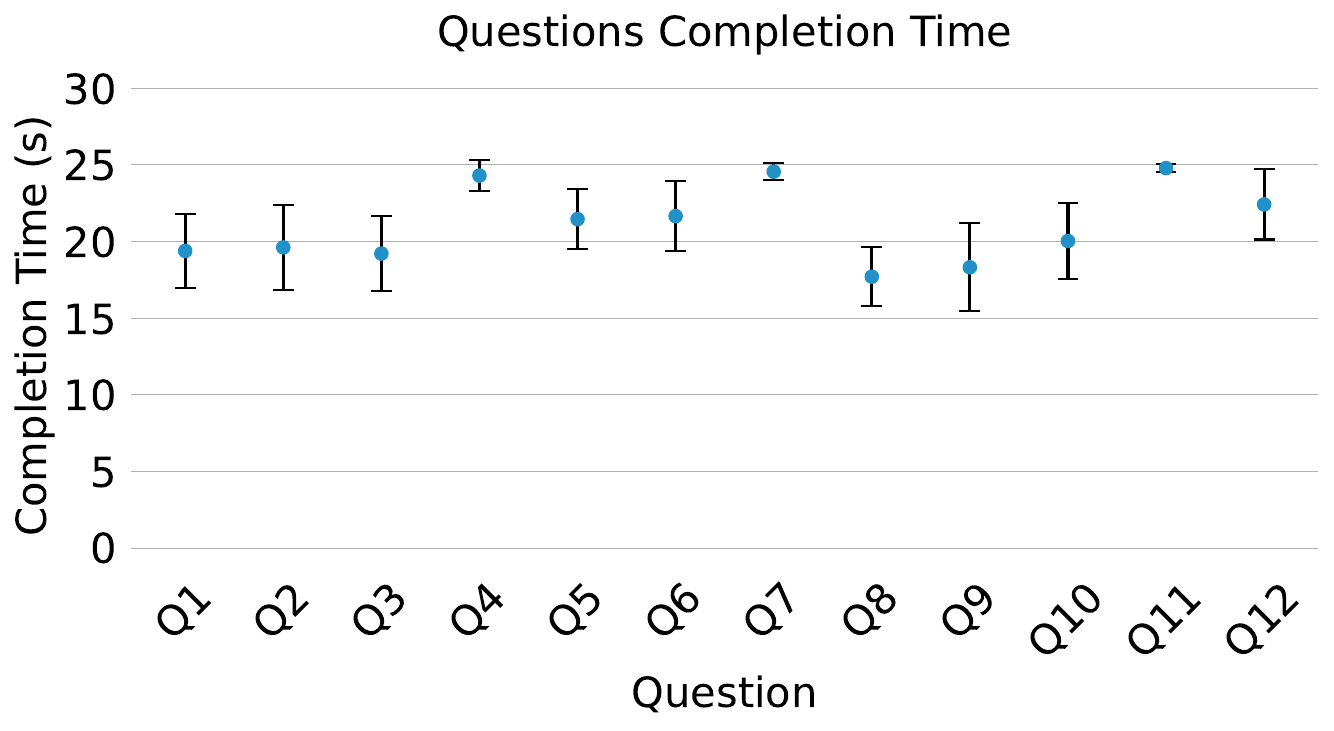}
    \caption{Completion time}
    \end{subfigure}
    \caption{Mini-VLAT test results with 13 participants from the architecture domain across 12 questions: (a) The number of correct, incorrect, and missing answers per question. (b) The completion time per question. The whiskers represent 95\% confidence intervals.}
    \label{fig:study_mini_vlat_result}
\end{figure*}
Our work is motivated by observations from a user study we conducted to assess the usability and usefulness of a visualization tool prototype~\cite{abdelaal2024visual} designed for domain experts in architecture. As a part of the evaluation procedure, we included the Mini-VLAT test in the introductory part of the study. The study participants had to perform the test before proceeding to the tasks using the tool. In total, we recruited 13 participants (five identified as females), and all had a university degree in architecture (seven master’s, five bachelor’s, and one Ph.D.). Before starting the test, we verbally explained the general structure of the test and the 25-second rule.

\autoref{fig:study_mini_vlat_result} shows the performance of the participants across the twelve questions. We observed behavior that indicated confusion and stress for the participants, which was also partially reported in other studies~\cite{NobreReading24}.
Our main observations were:
    \paragraph{\textbf{Stacked Charts Are the Most Challenging}} 
    Questions Q4, Q7, and Q11 stand out as the least correctly answered by participants. Q7 (``Stacked Bar Chart''), in particular, was answered correctly only once (\autoref{fig:study_mini_vlat_result_responses}). As highlighted in \autoref{fig:mini_vlat_questions_issues}, the question required multiple steps to identify correct colors, labels, etc., which was hard to perform within 25 seconds.
    A similar behavior can be noticed in Q11 (``Stacked Area Chart'').
    \paragraph{\textbf{Unfamiliar Choropleth Map.}}
     Most participants did not have a problem understanding question Q4 of whether Washington or Wisconsin has a higher unemployment rate or interpreting the visualization. However, they spent most of the time trying to locate the states on the map (see \autoref{fig:mini_vlat_questions_issues}). 
    \paragraph{\textbf{Line Chart and Bubble Chart Are the Easiest}} Questions Q6 (``Bubble Chart'') and Q8 (``Line Chart'') stand out as the easiest to solve. Most participants solved it correctly and in time. 
    \paragraph{\textbf{Trials Required}} Most participants needed to see the first question to fully understand the test format. Having one or two practice trials before starting the test would have been beneficial.
    \paragraph{\textbf{Confusion}} We observed that participants experienced confusion with some of the words or phrases used in the questions, labels (Q3), and colors (Q2). For Q9, it was difficult to estimate the correct answer due to small differences in the values (see \autoref{fig:mini_vlat_questions_issues}).
    \paragraph{\textbf{Sense of Skepticism}} We noticed a sense of skepticism among some participants as they struggled to understand the purpose of the test or how it relates to the visualization tool used later in the evaluation.
    \paragraph{\textbf{Stress and Time Pressure}} We observed that many participants felt stressed with a timer and some even lost focus when failing to answer on time.\vspace{2ex}

We did not have an open discussion with the participants to get their feedback on the test. Aside from the issues regarding phrasing, choice of colors, and axis labeling, these observations led us to question whether Mini-VLAT was the right format when working with domain experts.  Our rationale for collecting data on participants' visualization literacy was to potentially use it later to explain outliers, performance, and ratings in the results. 
While it is understandable that having to answer 12 questions within 25 seconds each was intended to scale up the test for a large number of participants, framing it as an assessment with time limits might have introduced other undesirable aspects. In particular, stress and time pressure are problematic for studies where performance in terms of completion times is not the focus. 
Although we informed the participants before the test that their performance would not impact the actual study and encouraged them not to feel pressured to answer all questions correctly, many seemed to experience stress due to the 25-second limit. 
During the actual user study with our visualization tool, we did not notice differences in the performance between participants who performed well in the test and those who did not.

While we think that the assessment of literacy was helpful in providing additional help, we wanted to investigate further how experts from different domains react to the test. On the one hand, to see if our observations could be repeated, and on the other hand, to find ways to improve the procedure so participants feel more comfortable performing the test.
\section{Structured Interviews}

\begin{figure*}[t]
    \centering   \includegraphics[width=.8\linewidth]{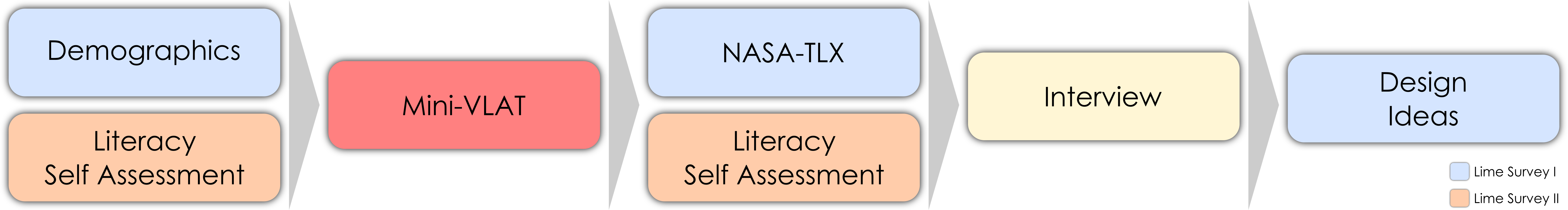}
    \caption{The study procedure consisted of a questionnaire using LimeSurvey and a structured interview. All participants had to perform the Mini-VLAT test before the interview.}
    \label{fig:experiment_procedure}
\end{figure*}

We conducted a study with experts from social science, architecture, engineering, and computer science. The study aimed to identify and address issues associated with the Mini-VLAT test. Participants were required to complete the test first to facilitate a comprehensive discussion and enhance participant empathy toward the test. 
We followed a mixed study design, including a questionnaire using LimeSurvey and a structured interview. \Cref{fig:experiment_procedure} depicts the procedure of the study.
We began by collecting demographic information and then proceeded with the Mini-VLAT test. After completing the test, participants assessed their workload using NASA-TLX. 
Furthermore, participants were asked to assess their visualization literacy before and after the test to determine the impact of their performance on their self-assessment.
The structured interview was conducted to discuss any issues encountered during the test in more detail. Finally, five different ideas for adapting Mini-VLAT were presented, and participants evaluated their usefulness compared to the original version of the test.

\subsection{Participants}

We invited seven scientific employees, identified as domain experts, to participate in our study. The participants comprised four females and three males, with five in the age group of 26--30 and two between 31 and 40.
The participants' professional backgrounds were from a wide range of 
disciplines from the methodological and application fields. They included two individuals from social science (\textbf{P1}, \textbf{P2}), two from computer science with visualization backgrounds covering methodological disciplines (\textbf{P3}, \textbf{P4}), two individuals from architecture with a design focus (\textbf{P5}, \textbf{P7}), and one from engineering  with a technical focus, covering disciplines from application fields (\textbf{P6}).
The participants reported their self-assessed experience level with data visualization. Three considered themselves slightly experienced (two from social science, one from architecture), two as moderately experienced (engineering, architecture), and two as experienced (computer science). 

\subsection{Interview}\label{sec:interview}
The structured interview was an important part of the study that critically examined different areas of Mini-VLAT. The interview was organized according to the following categories. The individual questions can be found in our supplementary material~\cite{darus}.
\paragraph{\textbf{Initial Impression of Mini-VLAT}}
After assessing the workload using the NASA TLX questions, we followed up by asking participants about their first impressions of the test and their feelings during and after the test.
\paragraph{\textbf{Question Type}}
We asked the participants about potential problems with the questions. Ambiguities in both the questions and visualizations were addressed, and the types of questions used were discussed. Participants were also asked to suggest changes that could improve the design of the test.

\paragraph{\textbf{Test Structure and Procedure}}
Firstly, they evaluated the overall length of the test and the effect of the time limit per question on their performance. We also investigated how the test environment, such as taking the test online from home, influenced their results. Finally, we inquired about the role of mental preparation, such as how prior information about the test's purpose might affect the performance.

\paragraph{\textbf{Performance Ranking}}
We asked participants to rank the specific factors (\textit{question type, test length, time limit for each question, test environment}) that they believed influenced their performance on the test. To facilitate this process, participants were given labeled cards representing each factor and were asked to arrange them according to their perceived impact on performance.

\paragraph{\textbf{Considering Different Target Groups}}
We considered the participants' views on using the test to measure visualization literacy in different target domains. 
Participants were asked to give their thoughts on using the test by domain experts from their own domain as well as from other target groups, including social scientists, visualization experts, architects, and engineers, and to indicate whether any group might be uncomfortable with the method or outperform others.
We also asked them for suggestions on adapting the test to make it more appealing and comfortable for these target groups.

\begin{figure*}[t]
    \centering
    \includegraphics[width=1\linewidth]{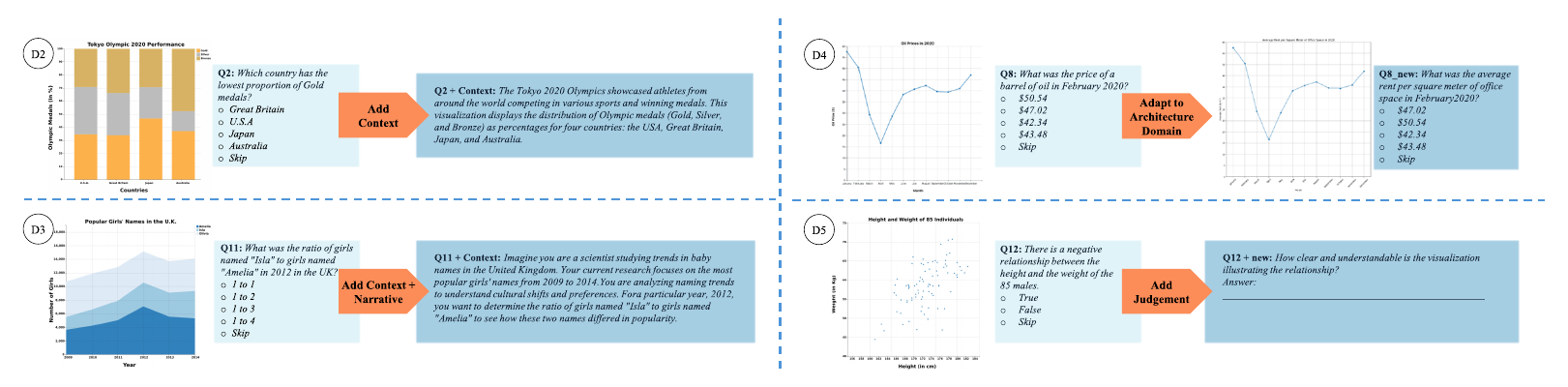}
    \caption{Individual questions from Mini-VLAT were applied to the defined design ideas (D2--D5). They were shown during the interview and evaluated in comparison to Mini-VLAT. Visualizations reprinted with permission from Pandey and Ottley~\cite{pandey_minivlat_2023} {\textcopyright} John Wiley \& Sons, Inc.}
    \label{fig:designIdeas}
\end{figure*}

\paragraph{\textbf{Alternative Design Ideas}}
In the final part of the interview, we presented our design ideas by examples (shown in \Cref{fig:designIdeas}) and asked participants to evaluate how they perceived our approach in comparison to  Mini-VLAT. Based on our findings from the previous study (\autoref{sec:preliminaryobservation}), we focused especially on making the questions easier to understand and providing a comfortable testing experience for the participant. We made the following five design adjustments:

    \begin{designNum}
        \item \textbf{Remove time limit.} 
        We proposed a test design without a time limit to create a more comfortable test environment and reduce errors caused by time pressure~\cite{slobounov2000timepressure}. It also served as an impulse to find out whether a time limit was desired. 
        \item \textbf{Add context to the question.} Since there is a correlation between cognitive features and visualization literacy~\cite{lee_correlation_2019}, we created a fitting context for a question in Mini-VLAT to improve the understanding of the question and the visualization.
        \item \textbf{Add context in a narrative way.} For another question, we presented the additional context in a narrative format to create a more engaging experience. Similar to Peck et al.~\cite{peck2019data}, where the familiarity of the visualization content captured participants' attention, we aimed for the narrative to help participants empathize with the situation.
        \item \textbf{Adapt to different domains.} We adapted a question from Mini-VLAT to the architecture domain without changing the visualization itself. The aim was to convey relevant content to the participants and catch their interest~\cite{peck2019data,hullman2018improving}.
        \item \textbf{Judge visualization.}  
        With this change, we aimed to let participants criticize the visualization themselves.
        Mistakes can result from misleading visualizations and questions~\cite{ge_calvi_2023}. In this way, participants could justify themselves.
    \end{designNum}
\begin{figure*}[t]
    \centering
    \begin{subfigure}{0.33\linewidth}
    \centering
       \includegraphics[width=\linewidth]{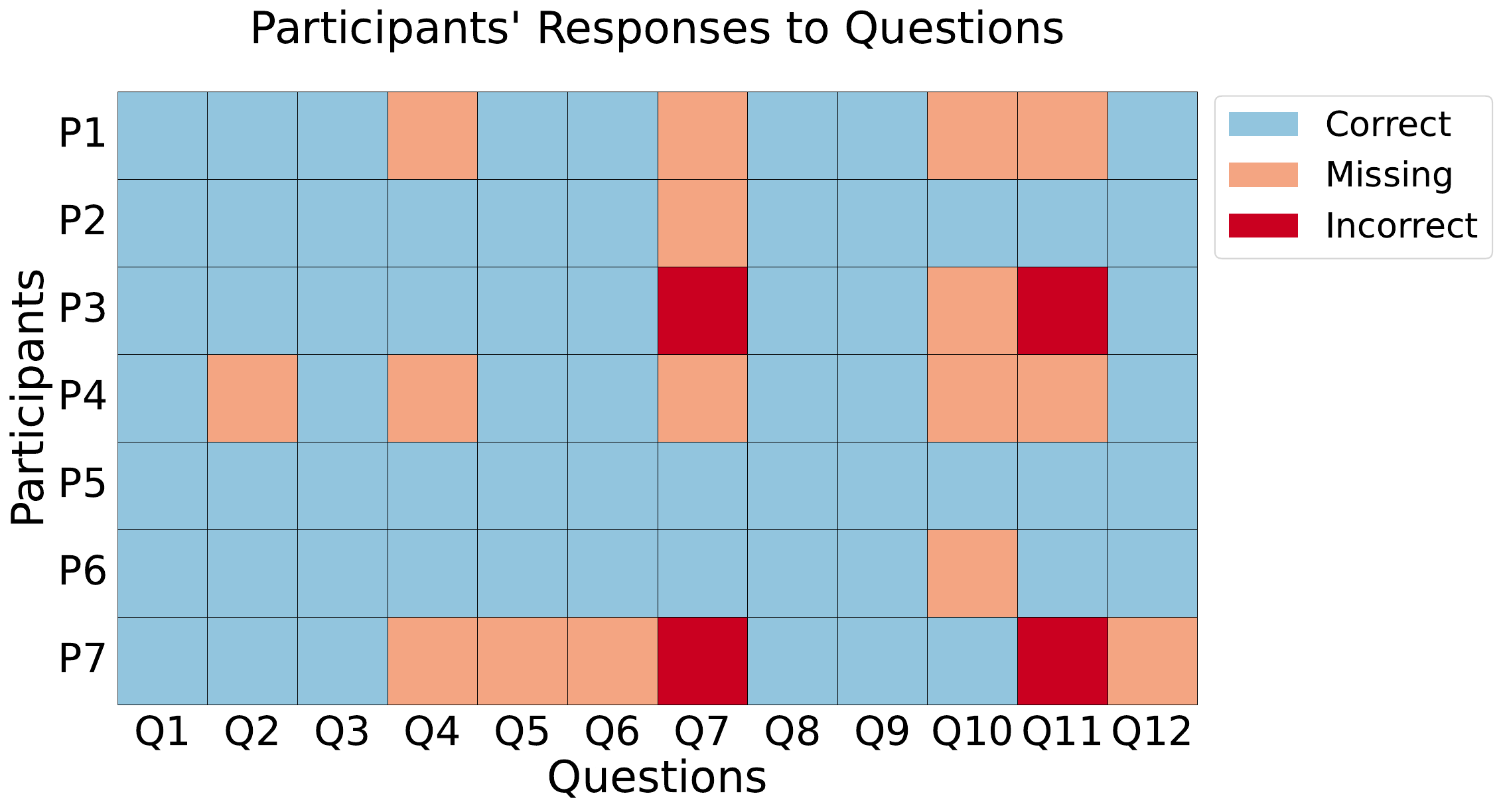}
    \caption{Mini-VLAT responses}
    \label{fig:participantsAnswersFreq}
    \end{subfigure}
    \begin{subfigure}{0.33\linewidth}
    \centering
    \includegraphics[width=.9\linewidth]    {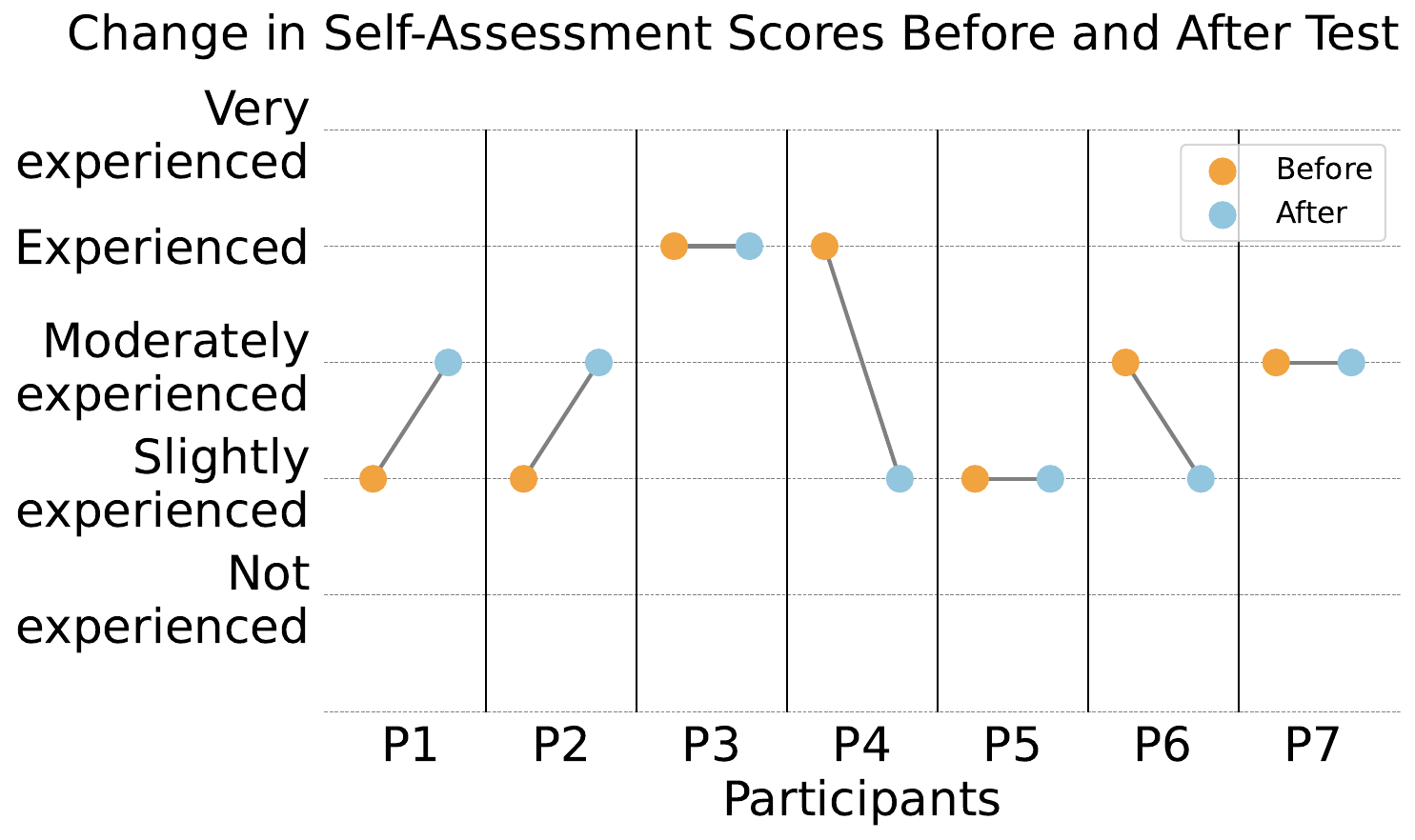}
    \caption{Self-assessment scores}
    \label{fig:selfAssessment}
    \end{subfigure}
    \begin{subfigure}{0.33\linewidth}
    \centering
    \includegraphics[width=.9\linewidth]    {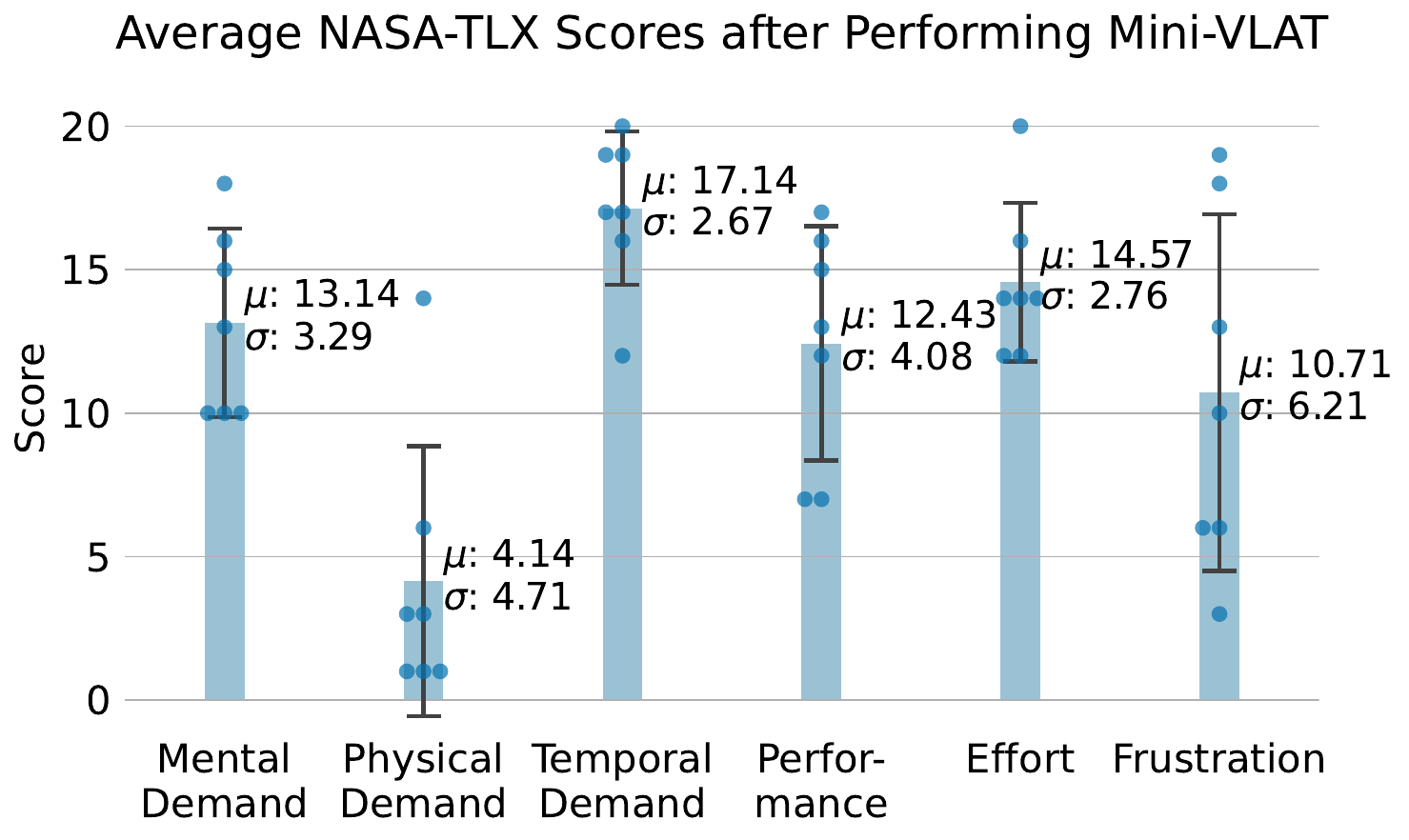}
    \caption{NASA-TLX results}        \label{fig:nasaTLX}
    \end{subfigure}
    \caption{Questionnaire results: (a) Participants' answers to the questions (Q1 to Q12) in the same order. Mistakes made by individual participants are marked as red. (b) Participants' subjective assessment of their visualization skills before~\protect\orangecircle~and after~\protect\bluecircle~the test. (c) Participants' perceived workload ratings using NASA-TLX after completing Mini-VLAT.}
    \label{fig:limesurvey_results}
\end{figure*}

\section{Results}
We recorded audio of the interviews and created automated transcripts from them using \textit{noScribe}.\footnote{\url{https://github.com/kaixxx/noScribe}, visited: August 30, 2024} For the analysis of the interviews, we applied Mayring's methodology for qualitative content analysis \cite{mayring2014qualitative}: We produced abstract summaries of the interview transcripts and systematically categorized the statements. Initially, we identified categories deductively by defining known issues and design ideas as individual categories. We derived this knowledge from our preliminary observations (\autoref{sec:preliminaryobservation}).
The remaining categories were identified inductively during the transcript analysis. The coded summaries of the individual interviews can be found in the supplementary material~\cite{darus}.
We grouped these categories under broader themes. In the following, we explore the issues and positive aspects identified in the interviews within these themes.

\subsection{Time}
As we noticed before, time is an important aspect of the test procedure. However, for some participants, it did not have only negative influences on the test experience.

\paragraph{\textbf{Time Limit}} Right at the beginning of the interview, we received various reactions from participants regarding the time limit. All participants reported that the time given was too short. 
As P4 stated, \textit{``It's hard to read, detect, assume, and give an answer in 25 seconds.''} This caused stress for some participants, as they were unable to complete their answers in time. 
Additionally, the time limit prevented participants from double-checking their answers. P3 stated: \textit{``I would
have answered some of the questions also with more confidence because sometimes I also want to double-check.''} After the test, they were concerned about which questions they had answered correctly. 

\paragraph{\textbf{Time Pressure}} The time constraint put all participants under pressure. \Cref{fig:participantsAnswersFreq} shows that three participants (P1, P4, P7) were noticeably bad at managing the time pressure. They could not answer at least four questions in time. P5 mentioned that the time pressure led to a loss of focus on the question: \textit{``And then I start looking at the time. How much time do I have left? Oh my God, I still haven't found where Wisconsin is. So that was the part that was bad for my focus.''}
As there were no breaks between questions, some participants were unable to focus on the next question after they failed to answer a question in time, which affected their performance. \Cref{fig:participantsAnswersFreq} illustrates this issue. Three participants were unable to answer or answered incorrectly after Q10. Specifically, P7 failed to answer three subsequent questions after Q4.

\paragraph{\textbf{Focus under Pressure}} In contrast, three participants (P2, P5, P6) handled the time pressure better than others. They performed relatively well (at least 91\% correct). They pointed out that the time limit positively affected them. The pressure helped them stay focused. As P5 noted, \textit{``I think sometimes it could be good. It keeps you focused with the pressure.''} P6 claimed that removing the time limit would reduce the challenge: \textit{``If there's no time limit at all, then (…) you're not as focused, (...) I'll just read it and look up and down and up and down again.''}

\paragraph{\textbf{Gamification}} \textit{``I found it quite fun to do; it was like gamification, like a game to see if you could grasp it in that short time.''} (P1). According to P1 and P2, the time limit added a playful effect when completing the test.

\subsection{Questions}
\label{sec:results_questions}

Regarding the test questions, we identified problems that partially also confirm findings from other studies~\cite{NobreReading24}.
\paragraph{\textbf{Ambiguities}} Participants observed some ambiguities in the questions. P7, for instance, got confused by Q12, which asked about ``men,'' who were not explicitly mentioned in the visualization:
 \textit{``I was thinking, maybe there are also girls, I don't know, like female and male. (...) It took me more time to realize if it was also in the graph.''}
P1 perceived the questions as trick questions and questioned the correctness of their answers. \textit{``The trick questions seem to be the most problematic to me because you quickly come into conflict with yourself.''}
Participants mentioned that a legend was misleading due to inconsistent ordering of the variables in the stacked bar charts, which caused frustration once recognized. For example, in the stacked bar chart of Q2 (see~\autoref{fig:mini_vlat_questions_issues}), the misleading color mapping to associate gold and bronze, along with the inconsistent order, became confusing for P6. \textit{``(...) gold is at the very bottom, and here it's at the very top. So I was like, which is gold, which is bronze because silver is obvious. (...) That was kind of hard.''}

\paragraph{\textbf{Complexity}} As P1 observed that different levels of complexity were present in the questions, several iterations of reading and viewing were necessary in order to answer these questions. \textit{``I first looked at the chart, and then I looked at the answers. And then I looked at the chart again.''} (P6).
Consequently, the time pressure was also greater for questions with higher complexity.

\paragraph{\textbf{Demographics}} The content of the questions relating specifically to the United States caused confusion among the participants. Q4 assumed general knowledge of locations of particular states in the USA, so P5 commented: \textit{``This is just not enough time unless I have a good grasp of geography. I'm not American. I don't know where Wisconsin is. That doesn't really make sense.''}

In the same way, the content of Q6 was described by P6 as confusing: \textit{``What's the price of an oil barrel? When I read the question, I was already thinking in my head, how much is an oil barrel in Germany? And I was like, oh, we don't measure that in barrels. And then I'm already a bit confused.''}

\paragraph{\textbf{Visualization}} 
Participants encountered difficulties in reading certain visualizations. \textit{``Visualizations generally include a caption, which was missing here.''} (P1). Two participants identified the challenge of reading values within the visualization due to the similarity in values across multiple attributes. P1 commented on this: \textit{``sometimes I had the feeling that it was really about a few pixels somehow, how high the bar is or something. I also found that a bit annoying when answering.''}

\subsection{Test Purpose} \label{sec:TestPurpose}
Many participants questioned the purpose of this test. It was unclear whether it was to assess general visualization skills or the level of visualization experience. Quoting P3, \textit{``The purpose of the test is probably just to check if someone has no experience at all, and then I think that's what you can figure out.''} P3 pointed out that other kinds of questions need to be asked to assess the experience level.

Also, P5 posed the question of whether visualization literacy means reading visualization or finding important information from visualizations. \textit{``Maybe the most interesting information about the chart isn't actually the question being asked. It would be nice to kind of contextualize it in the content.''} The questions should be formulated in such a way that the most relevant information is extracted from the visualization instead of asking about the reading of individual values.

Furthermore, P1 questioned the relevance of the time limit for the assessment, as it measures the efficiency of the interpretation and not the actual interpretation of the visualization in general. \textit{``As there is no time limit for the interpretation of visualizations in practice, it should not be necessary in the test either.''} (P1).

\subsection{Target Domains}
Our participants from different domains had different thoughts on the impact of the test on different target domains. Target groups who are generally used to taking tests would have fewer issues with the test.  P1 claimed that it is normal to feel uncomfortable in a test.
However, different target groups might perceive the test differently. Social scientists would think about the setting and have meta-thoughts. Architects might not be confronted with tests much in everyday life and could be overwhelmed by the tests. Visualization experts and engineers would not be bothered by it. P3 and P4 claimed as visualization experts that non-VIS people would feel more uncomfortable and overwhelmed with the test. 

In P5 and P6's view, domain experts from different target groups would be fine with the test because people are familiar with standardized tests. \textit{``That's also the way if you apply for a driver's license and things, that's always kind of test you have to do.''} (P6).
P5 believed that an important motivation for domain experts to take the test could be that it fulfills a meaningful purpose for their work.

\subsection{Self-Assessment}
As shown in \Cref{fig:participantsAnswersFreq} and \Cref{fig:selfAssessment},
the outcomes varied regarding their self-assessment scores and performance.
There were no major changes in the self-assessment scores. As the test results had not been communicated at this time, the participants could only guess their own performance. P4 recognized their poor performance (58\% correct), and their self-assessment score was much lower. However, the test caused P4 to question their abilities after underperforming. In the words of P4, \textit{``(...) in the end (...) I don't have that much experience because I would assume, like, an experienced visualization researcher (...) would be faster in graph interpretation.''} P7 also had a poor performance (50\% correct), but their self-assessment score remained unchanged for \mbox{unknown reasons.}

Participants assessed their workload using NASA TLX on a scale ranging from 0 (low) to 20 (high). \Cref{fig:nasaTLX} shows that the participants rated temporal demand the highest, which they also reported in the interviews. The participants had large differences in the frustration scores. Our observation was that some participants could better manage stress than others, which is reflected in \Cref{fig:participantsAnswersFreq}. 
On average, the participants made a high effort to perform the test. They also perceived high mental demand.

\subsection{Evaluating Design Ideas}

\begin{figure}
    \centering
    \includegraphics[width=\linewidth]{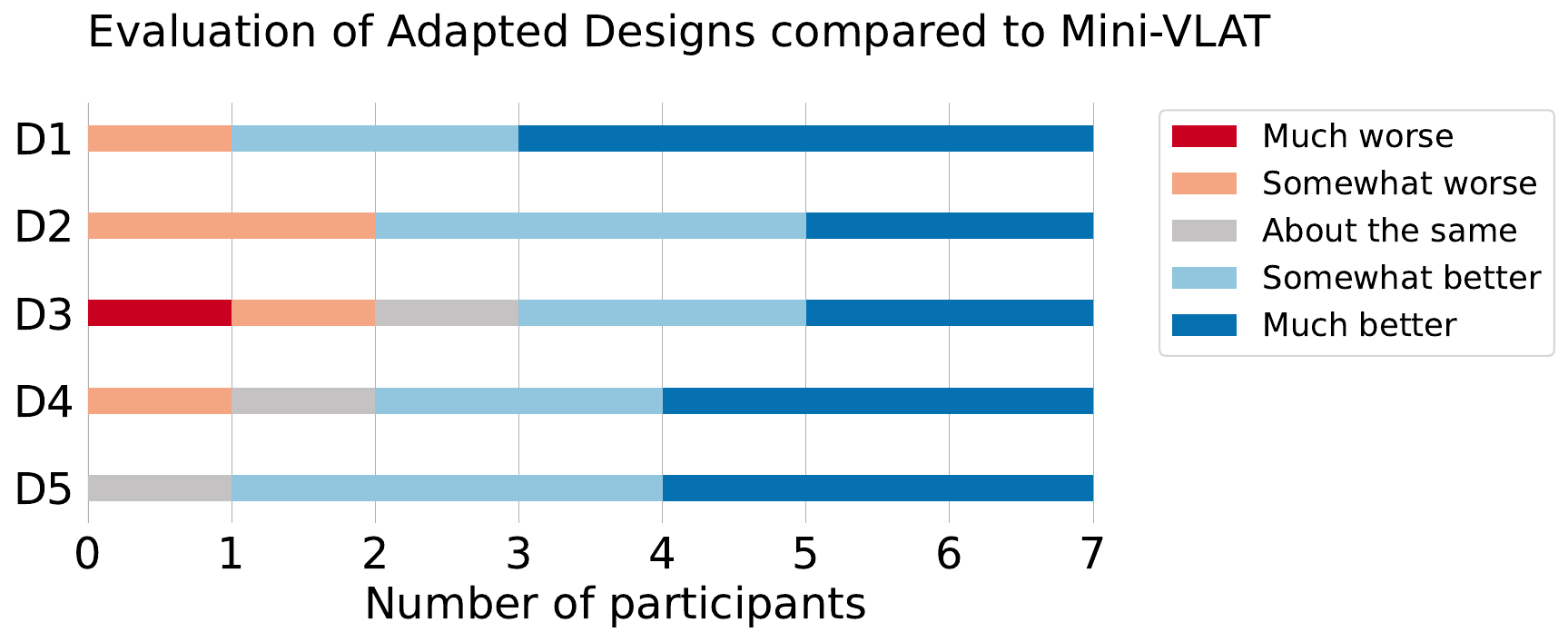}
    \caption{Participants' subjective ratings of 
    the presented design ideas.  
    D1 and D5 had the highest level of agreement.\vspace{-2ex}}
    \label{fig:designIdeaEvaluated}
\end{figure}

We discussed various ideas with our participants for adapting Mini-VLAT to make future assessments of visualization literacy more comfortable. They evaluated the proposed design ideas (\Cref{sec:interview}) and generally found them better than the original test, with only a few participants disagreeing (see \Cref{fig:designIdeaEvaluated}).\\

\paragraph{\textbf{Time Limit}} Almost all participants found it better to remove the time limit from Mini-VLAT (see \Cref{fig:designIdeaEvaluated}). P1 and P3 stated that having no time limit or at least more time would reduce errors and make participants more comfortable during the test. This would allow them to double-check answers and proceed with more confidence. Furthermore, they could not understand the reason behind the given time constraint and believed that such a limit was unnecessary for assessing the level of visualization literacy. 
Other suggestions regarding the time constraint included adapting the time limit to the complexity of the questions (P5). There was also the idea of having a time limit for the entire test, similar to exams at school (P4 and P5). 
In complete contrast, P6 rated that removing the time limit would be somewhat worse. According to P6, \textit{``I think it makes me a bit more focused, to be honest, because I know I only have 25 seconds and if I skip three questions, I won't perform as good.''}
P5 also shared this view. Removing the time limit would remove the challenge.

\paragraph{\textbf{Having Context}} Most participants found the addition of context in the question helpful (see \Cref{fig:designIdeaEvaluated}). P5 reasoned that having background information helps understand the purpose of the visualization. \textit{``It feels more natural, and it makes the activity feel more meaningful.''} 
On the other hand, the relevance of a context was questioned by P6. It would \textit{``extend[s] the process of finding an answer. It would cause more iterations: read, look, read, look….''} It would be superfluous to extend the test because the amount of information provided in the test was perceived as sufficient (P3, P7). However, people with less visualization experience could benefit more from the context (P4).

\paragraph{\textbf{Having Context + Narrative}}
There were a few differences in the rating compared to having only context. Most participants were positive about the change, but: \textit{``That makes the task more complicated. Because now I suddenly have to somehow understand a story and I'm not just being asked a fact.''} (P1). Since adding a context would only provide general information, that information would be more helpful. Similarly, participants P6 and P7, who found the adjustment worse, stated the narrative would require many more steps to understand the task.

\paragraph{\textbf{Domain Adaption}}
Five participants indicated that domain-specific questions are better than Mini-VLAT (see \Cref{fig:designIdeaEvaluated}). They stated that such questions would be more familiar and comfortable. P4 mentioned, \textit{``(...) For architects, prices and things they are doing all day. So, they have already some pre-knowledge about the data, then you don't have to think about it too much.''} This familiarity helps participants relate to the questions more and could influence their performance. P5, on the other hand, believed that this would not result in a significant change in performance.

Although the adaptation seems reasonable in comparison to the original test, P1 does not recommend such a change. \textit{``Without domain adaptation, it creates a distance between the self and the test. If the questions are tailored to your area of expertise and you fail the test, you will feel bad.''}

\paragraph{\textbf{Judging Visualization}}
Almost all participants liked the adaption of adding judgment of the visualization into the questions. As there were a couple of misleading visualizations in the test, one participant mentioned that this change would help provide feedback on the visualizations, allowing for improvements in later versions. 
P3 liked the change: \textit{``That's a great way to check the proficiency and experience level of someone. Because that takes a different kind of experience to not just answer the question, but actually understanding and also going beyond that, improving it. And that takes probably also more creativity.''}

Regarding perception, the questions would become more comfortable for the participants (P1), stating that \textit{``Perception is ultimately subjective, which means I can no longer make anything true or false, right or wrong.''}

\paragraph{\textbf{Test Procedure}} Besides the presented design ideas, participants also suggested further improvements for the test. One important aspect was an adaptation in the test procedure. Several participants (P1, P3, P4, P6) suggested including a training task to familiarize participants with the visualizations and question types beforehand. Additionally, P4 and P7 proposed having small breaks between questions, allowing participants to decide when to continue with the next question. These short breaks would help participants focus on the next question. 
P5 suggested using the time limit as a guideline rather than a strict rule and providing participants with information about the number of questions and the total time available during the test. It was also suggested that participants should be allowed to go back and revise previous questions, similar to an exam. P5 emphasized the importance of designing the test in an informal setting to provide a comfortable environment for domain experts.

\paragraph{\textbf{Question Type}} As mentioned in~\autoref{sec:TestPurpose}, P5 was concerned about the poor information content of the questions in the test. The questions should incorporate the main points of the visualization by adding more open-ended questions to focus on data interpretation rather than just reading.\vspace{2ex}

Additional suggestions included simplifying the selection of answers by allowing direct clicking on the relevant data within the visualization (P4). Another idea, proposed by P3, was to ask participants to identify a suitable visualization for a given dataset, either by sketching a visualization or by selecting one from a set of provided options.

\section{Discussion}

In the following, we discuss how the main issues identified in the interviews could be addressed. We further discuss use cases of how the test results can be used and considerations when working with domain experts.

\subsection{How Can We Improve Mini-VLAT?}
The interviews with domain experts confirmed our previous observations: There were several issues with Mini-VLAT. The test placed all participants under pressure due to time constraints. There were only a few participants who performed very well because they were able to manage stress and pressure successfully. Ambiguities and misreadings caused confusion in interpreting the questions and visualizations. The type of questions and the imposed time limit made participants question the overall purpose of the test.

\paragraph{\textbf{Reconsider the Time Limit}} The 25-second time limit in the original VLAT~\cite{lee_vlat_2017} was defined based on the average time crowd-sourced workers spent answering a question. In both of our studies, the allocated time was insufficient for most participants to answer most of the questions confidently. As crowd-sourced workers are used to taking tests, they might generally read and solve tasks faster~\cite{RossWho2010}. Also, questions have different complexities and, therefore, vary in the time required to answer. The time constraint pressured participants and made them question the relevance of the time limit. While some participants felt positively motivated by this, most of our participants reported that they provided incorrect or no answers to some of the questions due to time pressure. 
Accordingly, other studies on similar issues in the context of surveys on well-being~\cite{Fang2023} have shown that the quality of results increases when participants are given more time to reflect on individual questions. The authors~\cite{Fang2023} suggest implementing countdown timers in online surveys to prevent a question from being answered before a specific period.
However, they point out that respondents may find countdown timers stressful or distracting, which could also negatively affect their responses. 
Slobounov et al.~\cite{slobounov2000timepressure} investigated the effect of time pressure on performance and neurophysiological indices while performing a visuomotor task. The results showed that while task completion time under pressure was lower than without time pressure, the number of failed tasks increased.
The question of whether and how to implement time limits in assessing visualization literacy cannot be answered in general terms but must be decided on a case-by-case basis. It should not be forgotten that if the time limit is removed, time efficiency is not guaranteed. 
What seems to be generally agreed upon is that surveys should take as little time as possible~\cite{bhattacherjee2012social,sharma2022short} and provide as much time as \mbox{necessary~\cite{Fang2023}.}

\paragraph{\textbf{Revisiting the Visualizations}}
Some visualizations of Mini-VLAT tended to encourage misreading and misinterpretations, resulting in many participants answering questions incorrectly or not at all. This problem was also identified in the work of Nobre et al.~\cite{NobreReading24}.
We recommend revisiting all visualizations based on the issues identified in \autoref{fig:mini_vlat_questions_issues} and improving them accordingly.
Furthermore, many elements of the test consist of statistical graphics. Advanced techniques common in many visual analytics tools, such as SPLOMs or parallel coordinate plots, are not covered. If a separation between different levels of expertise is required, the test would require more variety in the types of visualizations.

\paragraph{\textbf{Reconsider the Questions Types}}
Some of the participants in the interview were confused by misleading questions. Similar to the changes to the visualizations, we also recommend revising ambiguous questions according to the findings in~\autoref{sec:results_questions}, which is consistent with previous work~\cite{NobreReading24}.
Most participants found adding context to the questions helpful, as it made them understand both the question and the visualization easily, making the experience more comfortable. 
As some participants indicated, the context can be particularly useful for those with less experience in visualization, but additional text requires more time to read and increases cognitive load, which can have a reverse effect on experienced people~\cite{kalyuga2007expertise}.
There were similar opinions regarding context + narrative. However, participants preferred the addition of context alone, as it ensures that only relevant information is read.
A recommendation in this context is not appropriate at this stage. Further research is needed to investigate the effect of adding context on participants' performance and perceptions at \mbox{different experience levels.}

\paragraph{\textbf{Adding Practice Trials and Time Breaks}}
In both studies, some participants became stressed due to missing breaks between questions in combination with the time limit. Participants wanted to move on to the next question when they were ready. Previous questions that caused stress affected the subsequent questions. With the help of fixed or user-defined breaks, participants could recover from the previous question and mentally prepare for the next question. 
Our studies also revealed the lack of a training task. Participants used the first task to familiarize themselves with the test. Similar to user studies, adding training tasks with simple visualizations would help participants better understand the test format and prepare them for the tasks. \\

We believe that a standardized visualization literacy test can be an efficient assessment method for diverse use cases, but we recommend revising Mini-VLAT with the proposed changes and ensuring its validity and reliability so that it can be used as a standardized assessment test. Moreover, future versions should publish the visualizations with a permissive license to ensure the reusability of the tests. This was not considered in Mini-VLAT and VLAT.

\subsection{Mini-VLAT Use Cases}
One important aspect to consider is the different scenarios and purposes for which Mini-VLAT can be applied. Based on our findings, we suggest the following use cases:

\paragraph{\textbf{An Instrument for Informing Design Studies}}
Design studies have a long tradition
in problem-driven visualization research. These studies involve visualization researchers working with domain experts to solve real-world problems, create visualization systems to address these issues, validate the designs, and refine guidelines based on their findings~\cite{sedlmair2012design}.
However, conducting design studies with and for domain experts poses several challenges that need to be considered. 
Immediate user involvement and sensitivity for the target domain are essential for their success~\cite{wortmeier2024configuring}. What has to be discovered and considered in an early stage are the specific practices, knowledge, competencies, and needs of end users \cite{sedlmair2012design}.
In this case, it is advisable to use Mini-VLAT less as a conclusive assessment of experts' competence and more as a basis for establishing a common ground between the visualization researchers and domain experts. In relation to the time limit of the test, it could be discussed beforehand whether decision-making under pressure and the ability to understand visualizations in a short time are requirements of the problem being addressed or whether certain types of questions and visualizations are relevant for the users.

\paragraph{\textbf{A Filtration Criterion in Crowd-sourced Evaluations}} Similar to HITs approval rate on mTurk, the results of Mini-VLAT can be used as a filtration criterion in online studies. In the simplest case, only people with sufficient visualization knowledge (whatever defined) are allowed to participate in the study.

\paragraph{\textbf{An Instrument for Interpreting Study Results}}
The results of the Mini-VLAT test can serve as an additional instrument to interpret the results of user studies. Similar to a demographics questionnaire, the test results can help explain some of the findings or form hypotheses regarding the participants' performance.

\paragraph{\textbf{A Training Criterion for Expert Studies}} 
Domain experts are typically hard to find, and in many evaluation scenarios such as expert studies~\cite{IsenbergSystematic2013}, leaving some out due to low levels of literacy might not be an option. In this case, the results of Mini-VLAT can be used not as a filtration criterion but rather as an indication that some experts might need additional examples, tutorials, and explanations, especially if the techniques under evaluation are more complex than the basic charts used in the Mini-VLAT test.

What remains open is the question of when the test should be applied in the study procedure. Low performance in the test might negatively influence the following tasks in a study. However, help and additional support based on test scores can only be provided if the test is performed in advance.
We further emphasize that good test scores do not necessarily guarantee a good performance in the rest of the study. As presented in the scenarios above, the interpretation of the results can have many different purposes.
Hence, authors should always state the purpose when applying such tests in their study procedures.

For the application in collaborative environments, such as workshops, we recommend handing out the test beforehand so participants can do it individually. Otherwise, it might create a competitive situation and distract participants from the real task.

\subsection{Considerations when Working with Domain Experts}

In the following, we outline considerations when assessing the visualization literacy of domain experts.

\paragraph{\textbf{Is a ``Test'' the Right Format?}}
When the goal is to assess the literacy of domain experts, the first question one should ask is \textit{if a test is the right format for assessment.} The test might lead to discontent among experts because it might put them in situations where they feel pressured because of time, exposed due to their lack of knowledge, or frustrated as they struggle to connect the competencies evaluated and their everyday work. 
In the social sciences, standardized tests are usually not used to examine the competencies of particular groups of people comprehensively, at least from the viewpoint of qualitative research. In that case, observation methods seem more appropriate for understanding experts’ skills, competencies, and know-how. In the visualization community, we see the work of Lee et al.~\cite{LeeKHLKY16} as one example of such work. Nevertheless, in the context of HCI and design~\cite{muller2018designethnografie, lazar2017research}, structured questionnaires and interviews are used to obtain quantitative insights, which could, in this case, also justify the use of Mini-VLAT.

\paragraph{\textbf{Stress-free Environment}}
 
When involving experts from specific areas, we seek their participation to support our research, and we cannot assume they will readily agree to take part in a typical test that induces pressure and stress. Therefore, it is crucial to create a comfortable test environment for them. This includes avoiding putting them under pressure---especially when the expert's ability to work under time pressure is not relevant to the evaluation. Furthermore, the test should be introduced in a simple and explanatory way so that the experts have time to understand what kind of questions are asked and how.

\paragraph{\textbf{Clear, Meaningful, and Relatable Questions}}

Studies on survey research projects show that many respondents find questions and measures often ambiguous, irrelevant, or misleading~\cite{einola2021behind}. We noticed an echo of these findings in our work (see~\autoref{sec:results_questions}). Questions should be designed so experts can read, understand, and respond to them in a meaningful way~\cite{bhattacherjee2012social}. This does not only mean clear wording adapted to the everyday and professional use of participants but also the practical relevance and cultural significance of the questions asked~\cite{rusche2023measuring}.
Our domain experts confirmed in the interview that the questions tailored to their specific domain make the questions more relatable and engaging. This is also consistent with previous findings~\cite{peck2019data,lee_correlation_2019}. The open question remains whether making the questions relatable, for example, by adapting them to the experts' domains or their personal experiences, might introduce a confirmation bias, as one of the social scientists pointed out.

\paragraph{\textbf{Culture Relevance}}
Questions with demographic and cultural references can also confuse participants and cause difficulties, as we found in our two studies when participants spent a long time searching for the states on the map. As stated by Einola and Alvesson~\cite{einola2021behind}, differences between cultures represent a significant obstacle to the reuse of questionnaires and similar instruments unless they are developed for and used on a very specific group during a limited time period. It should be considered that cultural differences not only occur between countries but also over time, between social classes, ethnic groups, professions, and organizations. What it means to be visually literate can vary from one specific group to another, even within the same domain~\cite{ahearn2004literacy,einola2021behind}. Therefore, when using the test, it is essential to adapt the questions to align with the everyday or professional knowledge of the specific target group. This may require prior exploration by the researcher.

\paragraph{\textbf{Ethical Considerations}}

The golden rule should always be followed to ensure a successful and ethical questioning process~\cite{bhattacherjee2012social}: \textit{``Do unto your respondents what you would have them do unto you. Be attentive and appreciative of respondents’ time, attention, trust, and confidentiality of personal information.''} It can, therefore, be added that participants should not only understand what kind of questions are to be answered in what way but should also be informed about why they are being asked.\\

Given the importance of collaboration with domain experts in visualization research, including design studies, developing a literacy test tailored to their needs would be a valuable contribution. However,
implementing a domain-adapted test that considers these aspects would not be fully generalizable. Compared to standardized tests with generalized assessment, its reusability with other domain experts would be limited. Further investigation into this approach is required.

\section{Conclusion}

Visualization literacy assessment is an important part of user-based evaluation procedures. However, this does not take into account the stress inflicted on the participants. Our results revealed potential issues related to using standardized tests in studies with domain experts. In an interview with experts, we identified that time constraints put participants under pressure. Ambiguities in the questions, and questions with unfamiliar demographic references led to confusion, which affected their test responses when taking the Mini-VLAT test. We discussed potential changes and provided suggestions for researchers who intend to assess their participants' visualization skills with Mini-VLAT. With the right framing and slight modifications of the test setup, visualization experts can find out if their participants require more help, e.g., by extended tutorials, while feeling more comfortable in the study situation.

We see some limitations of our findings in potential self-reporting biases and the fact that the expert domains did not include suggestions from fields related to educational methods and psychometrics.
We plan to expand to other research fields in future studies. We also plan to investigate our recommended design ideas further and develop a domain-adaptive test.

\acknowledgments{
This work is supported by the Deutsche Forschungsgemeinschaft
(DFG, German Research Foundation) under Germany’s Excellence
Strategy -- EXC 2120/1 -- 390831618.}
\bibliographystyle{abbrv-doi-hyperref-narrow}

\bibliography{beliv2024}
\end{document}